\documentclass{article}
\usepackage{amsmath}
\usepackage{psfrag}
\usepackage[dvips]{graphicx}
\advance\textheight by 45mm 
\newcommand{\muv}{\mu_0}
\newcommand{\Z}{z}
\newcommand{\D}{\,\mathrm{d}}
\newcommand{\avrg}[1]{\ensuremath{\left<#1\right>}}
\newcommand{\pd}[2]{\ensuremath{\frac{\partial #1}{\partial #2}}}
\newcommand{\metric}{\ensuremath{g^{\alpha\beta}}}
\newcommand{\metricC}{\ensuremath{g_{\alpha \beta}}}
\newcommand{\ie}{\textit{i.e.,}~}
\newcommand{\eg}{\textit{e.g.,}~}

\begin{document}
\psfragscanon
\author{
A.F.\,Andreev\footnote{E-mail: andreev@kapitza.ras.ru},
L.A.\,Melnikovsky\footnote{E-mail: leva@kapitza.ras.ru}
}
\title{Two-velocity elasticity theory and facet growth}
\date{}
\maketitle
\begin{center}
\textit{Kapitza Institute for Physical Problems}\\
\textit{117334, Moscow, Russia}\\
\end{center}

\begin{abstract}
We explain the linear growth of smooth solid helium facets by the presence of
lattice point defects. To implement this task, the framework of very general
two-velocity elasticity theory equations is developed. Boundary conditions
for these equations for various surface types  are derived. We also suggest
additional experiments to justify the concept.
\end{abstract}


\section{Introduction}

\begin{figure}[hb]
\begin{center}
\includegraphics[scale=1]{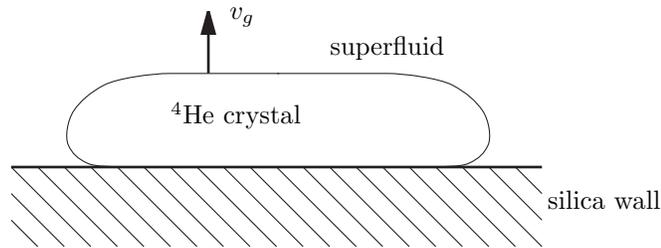}
\end{center}
\caption{Typical experimental layout} \label{sketch}
\end{figure}

Existence of two distinct states of a crystal surface is well known: it may
be either  {\em smooth} or {\em rough} (for a review, see~\cite{nozieres}). A
smooth surface is characterized by a long-range order and small fluctuations.
On the contrary, a rough surface behaves differently~--- it does not exhibit
a long-range order and its displacement fluctuates heavily. These equilibrium
properties lead to different kinetic properties. While the rough surface is
usually supposed to grow easily (as described by the growth coefficient), the
smooth one (providing the crystal has no dislocations) is characterized by
zero growth coefficient and grows with nuclei of new atomic layer. In
accordance with this mechanism, one should not observe the linear growth rate
if low overpressure is applied. Reality is different~---
experiments~\cite{exp} demonstrate that a smooth helium surface free of screw
dislocations grows linearly. This work is an attempt to explain this behavior
by the presence of lattice point defects (vacancies). The idea is similar to
that suggested by Herring~\cite{herring} and by Lifshitz~\cite{iml} as an
explanation of polycrystal flow. It is quite simple: the mass flux in bulk
helium is attributed to the motion of vacancies. This flux is the mass
transfer through the lattice. Therefore, if vacancies are allowed to be
created on the bottom edge of the sample (the boundary between the crystal
and the wall, see Fig.~\ref{sketch}) and to annihilate on the top of it (on
the smooth crystal-liquid interface), then the crystal grows.

The suggested crystal growth mechanism may be explained as follows. Since the
smooth crystal facet (the top one in Fig.~\ref{sketch}) cannot move with
respect to the lattice, it moves upward stuck to the lattice. Vacuities to
appear due to this on the bottom edge of the crystal transform into lattice
defects (ordinary vacancies) and go up through (and faster than) the bulk
helium. They finally vanish in the liquid on the top smooth surface of the
sample. In other words, the crystal grows on the boundary between helium and
the wall, rather than on the smooth solid--liquid interface (which
nevertheless provides mass supply for the growth). It is important to
emphasize that this scenario may occur if and only if the vacancies are
allowed to emerge on the bottom edge of the crystal. One may say that this
boundary is in some sense ``atomically rough''~--- it can grow new atomic
layers. For this condition to be satisfied, the wall surface should have a
disordered shape or be slightly tilted with respect to basal planes of the
crystal (thereby forming a vicinal interface). This ensures that the surface
can play the role of a source or a sink of vacancies. An atomically flat wall
parallel to the basal plane should, in contrast, behave like a normal smooth
surface~--- it is fixed to the lattice. This is due to the fact that for the
surface to move new atomic layer nuclei have to be created.

This paper is organized as follows. In Secs.~\ref{section_defs} and
\ref{section_eq_and_fluxes} we derive very general two-velocity elasticity
theory equations. They consist of conditions for the conventional elasticity
theory variables (including lattice velocity) and equations for a macroscopic
description of the quasiparticle gas (including the quasiparticle gas
velocity).

Equations to be derived are similar to those of the two-velocity superfluid
hydrodynamics. Velocities of lattice and excitation gas in our equations
replace superfluid and normal component velocities of the two-fluid
hydrodynamics. Similarly to the usual linear phonon hydrodynamics
(see~\cite{LL10} \S~71) Umklapp-process (which result in the
nonconservation of the total quasimomentum in quasiparticle collisions)
probability is supposed to be low. In the low-temperature region considered
here this assumption is quite reasonable. We also neglect dissipation here.
This means that our analysis is limited to the terms of the first order in
gradients.

Derivation procedure of exact (nonlinear) hydrodynamics equations for
superfluid can be implemented (see~\cite{khalat}) from phenomenological
considerations, using conservation laws. A constitutive argument for this
derivation is the statement that the superfluid flow is potential. This is an
intrinsic property of the order parameter in a superfluid. Such condition is
unavailable for a crystal (moreover, there is no quasimomentum conservation
relation in the nonlinear description, see Eq.~\eqref{dotP} below).

We deduce the two-velocity elasticity theory equations using a more general
approach (see the paper by Pushkarov and one of the authors~\cite{pushka}, as
well as~\cite{abazal} and~\cite{abs}). It is based on the kinetic equation
description of the quasiparticle dynamics. The realization of this technique,
particularly in nonlinear situation, is a matter of
considerable interest for its own sake not only for a solid but also for a
superfluid (that this procedure is possible is mentioned in~\cite{LL10}
\S\,77). With this technique, we find exact expressions for all hydrodynamic
variables and their dependence (in terms of the quasiparticle energy
spectrum) on the relative velocity of components. It is trivial to extend the
equations obtained for the solid dynamics to the simpler case of superfluid
hydrodynamics.

Boundary conditions for our equations depend on the surface type; in
Sec.\ref{section_bound_cond} we thoroughly consider three possibilities:
rough (Sec.~\ref{section_sl_rough}) and smooth (Sec.~\ref{section_sl_smooth})
interfaces between solid and liquid helium, and the rough boundary between
solid helium and normal hard wall (Sec.~\ref{section_sw_rough}). Finally, in
Sec.~\ref{section_growth} we calculate the growth rate for the crystal.

\section{Definitions}
 \label{section_defs}
Following  the principles in Refs.\cite{pushka,abazal,abs} we employ the
Euler approach to the lattice description. We thus introduce three ``node
numbers'' $N^\alpha$ ($\alpha=1,2,3$). They are functions of space
coordinates $\mathbf{r}$ and time $t$, $N^\alpha = N^\alpha(\mathbf{r},t)$.
From now on, Greek indices (like $\alpha$ here) are used for the ``lattice
space'' and Latin indices (\eg $i$ in $x_i$ for the components of
$\mathbf{r}$) for the real space components. Defining the reciprocal lattice
vectors as
 $\mathbf{a}^\alpha=\partial N^\alpha /\partial \mathbf{r}$,
we get the elementary lattice translation vectors $\mathbf{a}_\beta$ as
 $\mathbf{a}^\alpha \mathbf{a}_{\beta}=\delta^{\alpha}_{\beta}$.
Taking the time derivative, we obtain the lattice velocity as
$\mathbf{w}=-\mathbf{a}_\alpha \dot{N}^\alpha$. The elastic energy $E_l$ of
the lattice is a function of the deformation. Moreover, since it depends not
on the spatial orientation of the infinitesimal sample (the space is
isotropic), but on the relative position of the $a^\alpha_i$ vectors, we may
write $E_l=E_l\left(\metric \right)$, where $\metric =\mathbf{a}^\alpha
\mathbf{a}^\beta$ is a symmetric ``metric tensor'' of the lattice space.

We are now ready to describe quasiparticle degrees of freedom. We do not
specify the quasiparticle nature at the moment (be it phonons, vacancies as
in~\cite{abs}, or electrons as in~\cite{pushka}). Actually, all the equations
below imply the summation over all branches of excitations; we do not
explicitly write the sum for brevity. Any quasiparticle should be
characterized by its mass $m$ (zero for phonons, positive for electrons, and
negative for vacancies), coordinate, and momentum. Since quasiparticles exist
in the lattice background, the quasimomentum should be used. Their energy in
the frame of reference of the lattice $\epsilon=\epsilon(\mathbf{a}_\alpha
(\mathbf{p} - m \mathbf{w}), \metric )$ is a periodic function of the
quasimomentum $\mathbf{p}$ (its periods are $2\pi\hbar \mathbf{a}^\beta$). In
laboratory frame of reference, we have the quasiparticle energy
(see~\cite{pushka})
\[
\tilde{\epsilon}=\epsilon+m\mathbf{w}\pd{\epsilon}{\mathbf{p}}+m\frac{w^2}{2}.
\]
We also use the variables $\mathbf{k} = \mathbf{p} - m \mathbf{w}$ and
$k_\alpha=\mathbf{a}_\alpha \mathbf{k}$. Quasiparticle dynamics is determined
by the Hamilton function
\[
H=\epsilon+\mathbf{p w} - mw^2/2.
\]
We now introduce the distribution function $f(\mathbf{r},\mathbf{p})$ (it is
also a periodic function of the quasimomentum $\mathbf{p}$). Its kinetics is
governed by the Boltzmann equation
\begin{equation}\label{Boltzmann}
\pd{f}{t} + \pd{f}{\mathbf{r}} \pd{H}{\mathbf{p}}- \pd{f}{\mathbf{p}}
\pd{H}{\mathbf{r}}= \mathrm{St} f.
\end{equation}
Using this distribution function, we can obtain macroscopic quantities like
the mass density as
\[
\rho= \rho_l + \avrg{m f} = M \det \left( \metric \right)^{1/2}+m n,
\]
where the angle brackets denote the integration over quasimomentum space,
$\avrg{\phantom{|}}=\int\D^3 p/(2\pi \hbar)^3$, $\rho_l$ is the lattice
density, $M$ is the mass of an elementary cell, and $n=\avrg{f}$.

We consider a quasi-equilibrium distribution function. The complete set of
quantities conserved in quasiparticle collisions consists of their mass
(proportional to their quantity for ``real'' particles like electrons and
vacancies and zero for phonons), energy, and quasimomentum (in the
low-temperature region,  Umklapp processes may be neglected). Consequently,
the most general quasi-equilibrium distribution is a function of
\[
\Z
=\frac{\epsilon-\mathbf{kv}-m\muv}{T}=\frac{\epsilon-(\mathbf{p}-m\mathbf{w})\mathbf{v}
-m\muv}{T}=
\frac{\epsilon-\mathbf{pv}-m\muv+m\mathbf{wv}}{T}=\frac{\epsilon-\mathbf{pv}-m\phi}{T},
\]
where $\muv-\mathbf{wv}=\phi$. The Lagrange coefficients $T$, $v$, and $\muv$
denote temperature, the velocity relative to the lattice, and chemical
potential of the quasiparticle gas, accordingly. For definiteness, we assume
that the excitations are Bose particles. The distribution function is then
given by
\begin{equation}
\label{distribution} f=\frac{1}{e^ \Z
-1}=\left(\exp\frac{\epsilon-\mathbf{pv}-m\phi}{T}-1\right)^{-1},
\end{equation}
and $\ln ((f+1)/f)= \Z $.\footnote{For further convenience, we also provide
here the result of the distribution function integration:
\[
\int f \D  \Z  = \ln \frac{e^ \Z -1}{e^ \Z } = - \ln (f+1).
\]}

We can now calculate other macroscopic parameters with this distribution
function. For the mass flux, we have
\begin{equation}\label{massflux}
\mathbf{J}=\rho_l \mathbf{w} + \mathbf{w} m n + \mathbf{j} = \rho \mathbf{w}
+ \mathbf{j} =
   \rho \mathbf{w} +  mn \mathbf{v},
\end{equation}
where the mass flux with reference to the lattice is
\[
\mathbf{j}=m\avrg{f\pd{\epsilon}{\mathbf{p}}}=mn \mathbf{v}.
\]
Using $\mathbf{J}$, we can write the mass conservation as
\[
\dot{\rho}+J_{i,i}=0.
\]
The number of real (massive) particles is also conserved in the bulk, and we
therefore have one additional conservation law
\[
\dot{\rho_l}+(\rho_l w_i)_{,i}=0.
\]

Similarly, the energy density is given by
\begin{equation}
\label{E_microscopic} E=\rho_l\frac{w^2}{2}+E_l\left(\metric
\right)+\avrg{\tilde{\epsilon}f}=
 \rho\frac{w^2}{2}+E_l\left(\metric \right)+\mathbf{wj}+
 \avrg{\epsilon f}.
\end{equation}
This equation allows us to prove (and find) exact macroscopic equivalents of
the microscopic quantities introduced above. The total energy density of the
crystal can be obtained via a Galilean transformation,
\begin{equation}
\label{E_macroscopic} E=E_0 + \frac{\rho w^2}{2} + \mathbf{j w},
\end{equation}
where $E_0=E_0(\mathbf{a}^\alpha, S, \rho, \mathbf{K})$ is the energy in the
frame of reference of the lattice, with $\mathbf{K}=\avrg{\mathbf{k} f}$
characterizing the quasimomentum density. A reasonable expression for the
$E_0$ differential
\begin{equation}
\label{E0_macroscopic}\D E_0 = \lambda_{ij} a_{\nu j}\D a^{\nu}_i + T \D S +
\mu \D \rho + \mathbf{v} \D \mathbf{K}
\end{equation}
we obtain with the conventional definition of the entropy density for the
Bose gas,
\[
S=\avrg{(f+1)\ln (f+1) -f \ln f} = \avrg{fx + \ln (f+1)}.
\]
Its differential being
\[
\D S=\avrg{(\ln (f+1) +1 - \ln f -1)\D f} = \avrg{ \Z  \D f}.
\]
Subtracting the differentials of~\eqref{E_microscopic}
and~\eqref{E_macroscopic}, we obtain
\begin{multline}
\label{EminusE1}
 0=
  \lambda_{ij} a_{\nu j}\D a^{\nu}_i
  + T \D S + \mu \D \rho +  \mathbf{v} \D \mathbf{K}
  -\D E_l\left(\metric \right)-\D \avrg{\epsilon f}\\ =
\lambda_{ij} a_{\nu j}\D a^{\nu}_i
  +  T \avrg{ \Z  \D f} + \mu \D \rho +  \mathbf{v}\D \mathbf{K}
  - \D E_l\left(\metric \right)- \avrg{\epsilon \D f}- \avrg{f \D \epsilon }\\ =
\lambda_{ij} a_{\nu j}\D a^{\nu}_i
  +  \avrg{(T  \Z  - \epsilon +  \mathbf{v k})\D f} + \mu \D \rho
  - \D E_l\left(\metric \right)
    +  \mathbf{v} \avrg{ f \D  \mathbf{k}} - \avrg{f \D \epsilon }
\end{multline}
We now transform the part of this equation related to the lattice deformation
\begin{multline*}
   \frac{\mu}{2}\rho_l \metricC \D\metric
 - \D E_l\left(\metric \right) +  \mathbf{v} \avrg{ f \D  \mathbf{k}}
 - \avrg{f \D \epsilon }=
  \frac{\mu}{2}\rho_l \metricC \D\metric
 - \pd{E_l}{\metric }\D\metric  - m \mathbf{v} \avrg{ f \D  \mathbf{w}}
 \\ -
 \avrg{f \left(
           \pd{\epsilon}{k_\alpha} \left(
                                      \left(\mathbf{p}-m\mathbf{w}\right)
                                      \D \mathbf{a}_\alpha
                                      - m \mathbf{a}_\alpha \D
                                      \mathbf{w}
                                     \right)
            +\left(\pd{\epsilon}{\metric }\right)_{k_\alpha}\D\metric
          \right) }\\ =
   -\left(
       \pd{E_l}{\metric }
       +\avrg{f \left(\pd{\epsilon}{\metric }\right)_{k_\alpha}}
   \right) \D\metric
   +\left( \mu \rho_l \delta_{ij}
          -\avrg{f \pd{(T x+\mathbf{p v})}{p_j} \left(p_i-mw_i\right)} a^\alpha_j
    \right)\D a_{\alpha i}\\ =
   -\left(
        \pd{E_l}{\metric }
       +\avrg{f \left(\pd{\epsilon}{\metric }\right)_{k_\alpha}}
   \right) 2 a^{\alpha}_i a^{\beta}_j a_{\nu j} \D a^{\nu}_i\\
   +\left( \delta_{ij} (T  \avrg{\ln(f+1)} + \mu \rho_l)
      +v_i P_j -m n v_i w_j
   \right) a_{\nu j} \D a^\nu_i\\ =
      \left(
            \delta_{ij} \left(T S + \rho\frac{w^2}{2}+E_l
            -  E + \mathbf{P v} + \mu \rho_l + mn\muv
      \right)
      -\Lambda_{ij}
      +v_i P_j -m n v_i w_j
   \right) a_{\nu j} \D a^\nu_i
.
\end{multline*}
Finally, from \eqref{EminusE1} we get:
\begin{multline}
\label{EminusE2}
0= (\mu - \muv) m \D n\\
    +\left\{
    \lambda_{ij}
    -\Lambda_{ij}
   + \delta_{ij} \left(T S +E_l - E + \mu \rho + \mathbf{P v}+ \rho\frac{w^2}{2} \right)
   +v_i P_j -m n v_i w_j
     \right\}
   a_{\nu j} \D a^\nu_i
,
\end{multline}
where we introduced $P_i=\avrg{p_i f}$ and
\begin{equation}
\label{Lambda}
 \Lambda_{ij}= 2 a^{\alpha}_i a^{\beta}_j
 \left(\pd{E_l}{\metric }
       +\avrg{f \left(\pd{\epsilon}{\metric }\right)_{k_\alpha}}
   \right).
\end{equation}
The terms in~\eqref{EminusE2} are independent, and each of them must
therefore be equal to zero. That is,
\[
\mu=\muv,
\]
\begin{equation}
\label{lambda_ij}
 \lambda_{ij} = \Lambda_{ij}
   - \delta_{ij} \left(TS +E_l - E + \mu \rho + \mathbf{Pv}+ \rho\frac{w^2}{2} \right)
   -v_i P_j + mn v_i w_j.
\end{equation}

\section{Equations and Fluxes}
\label{section_eq_and_fluxes}
 Here, we derive dynamics equations and thermodynamic fluxes for the system.
Neglecting dissipation at this point, we assume that the entropy conservation
law is valid,
\[
\dot{S}+F_{i,i}=0,
\]
where the entropy flux $F_i$ is determined by
\[
\mathbf{F}=S(\mathbf{v}+\mathbf{w}).
\]

We continue with the  equation for the momentum flux found in~\cite{pushka},
\begin{equation}
\label{dotJ}
 \dot{J_i}+\Pi_{ik,k}=0,
\end{equation}
where
\begin{multline}
\label{momentum_flux}
 \Pi_{ik}=\rho w_i w_k -
         E_l \delta_{ik} +
         2a^{\alpha}_{i}a^{\beta}_{k}
            \left(\pd{E_l}{\metric } +
                  \avrg{f
                  \left(\pd{\epsilon}{\metric }\right)_{k^\alpha}}
            \right)
         +w_i j_k + w_k j_i \\ =
   \rho w_i w_k -
   E_l \delta_{ik} +
   m n(w_i v_k + w_k v_i)+
            2a^{\alpha}_{i}a^{\beta}_{k}
            \left(\pd{E_l}{\metric } +
                  \avrg{f
                  \left(\pd{\epsilon}{\metric }\right)_{k^\alpha}}
            \right)
                    \\ =
   \rho w_i w_k -
   E_l \delta_{ik} +
   m n(w_i v_k + w_k v_i)+
         \Lambda_{ik}
 ,
\end{multline}
where we employed the definition~\eqref{Lambda} for~$\Lambda_{ik}$.

Taking the appropriate equation for the energy flux from~\cite{pushka}, we
have
\begin{equation}
\label{dotE} \dot{E}+Q_{i,i}=0,
\end{equation}
where
\[
Q_i=w_i E_l + \avrg{\epsilon \pd{H}{p_i} f } - \frac{w^2}{2}J_i +
    w_k\Pi_{ik}.
\]
To find the second term, we again use the distribution function
from~\eqref{distribution},
\begin{multline*}
 \avrg{\epsilon \pd{H}{p_i} f } =
   \avrg{\epsilon \left(\pd{\epsilon}{p_i} +w \right) f } \\ =
   \avrg{\epsilon \left(T \pd{ \Z }{p_i} + v_i + w_i \right) f }  =
   (v_i + w_i)\avrg{\epsilon f }
   + T \avrg{( T x + \mathbf{p v} + \phi) \pd{ \Z }{p_i} f }\\
    =
   (v_i + w_i)\avrg{\epsilon f } - T \avrg{ \mathbf{p v}  \pd{\ln(1+f)}{p_i}}  =
   (v_i + w_i)\avrg{\epsilon f } + T v_i \avrg{ \ln(1+f) } \\ =
   (v_i + w_i)\avrg{\epsilon f } +
     T v_i \avrg{ \ln(1+f) + f\ln\left(\frac{f+1}{f}\right) } -
     T v_i \avrg{            f\frac{\epsilon-\mathbf{p v}-m\phi}{T}}  \\ =
   (v_i + w_i)\avrg{\epsilon f } +
     T v_i S - v_i \avrg{ f\epsilon} + v_i \mathbf{v P} + mn\phi =
   w_i\avrg{\epsilon f } + v_i (T S + \mathbf{v P} + mn\phi)
.
\end{multline*}
For the energy flux, we finally have
\begin{multline}
\label{energy_flux}
 Q_i=
  w_i\avrg{\epsilon f } + v_i (T S + \mathbf{v P} + mn\phi)
  - \frac{w^2}{2}(\rho w_i +  m n v_i)
  +  w_k( \rho w_i w_k + m n(w_i v_k + w_k v_i) + \Lambda_{ik})\\ =
  w_i\avrg{\epsilon f } +
  v_i (T S + \mathbf{v P} + mn\phi) +
  \frac{w^2}{2}(\rho w_i + m n v_i)
  +  w_k( m n w_i v_k  + \Lambda_{ik})\\ =
  w_i \left(\avrg{\epsilon f } + \mathbf{w j} + \rho\frac{w^2}{2} \right)+
  v_i \left(T S + \mathbf{v P} + m n \left(\phi + \frac{w^2}{2}\right) \right) +
  w_k \Lambda_{ik}\\ =
  v_i \left(T S + \mathbf{v P} + m n \left(\phi + \frac{w^2}{2}\right) \right) +
  w_i (E - E_l)+
  w_k \Lambda_{ik}
.
\end{multline}
This formula completes the list of the conventional elasticity theory
equations. An additional equation is required to govern the quasiparticle
degrees of freedom. We now find the time derivative of $P_i$. We multiply
Boltzmann equation~\eqref{Boltzmann} by $p_i$ and integrate over the momentum
space. We temporarily neglect Umklapp processes, which are supposedly rare.
If needed, dissipation can be explicitly introduced into the final result. In
other words, the quasimomentum $\mathbf{p}$ is conserved in (normal)
collisions, and the term involving the collision integral is therefore zero.
The left-hand side of the Boltzmann equation gives
\begin{multline*}
\dot{P_i}=\avrg{p_i\left(\pd{f}{\mathbf{p}}
                         \pd{H}{\mathbf{r}}-
                         \pd{f}{\mathbf{r}}
                         \pd{H}{\mathbf{p}}\right)} \\=
\avrg{p_i\left(\pd{f}{\mathbf{p}}
         \left(\pd{\epsilon}{\mathbf{r}}+
              (p_j-mw_j)\pd{w_j}{\mathbf{r}}\right)-
               \pd{f}{\mathbf{r}}
         \left(\pd{\epsilon}{\mathbf{p}}+
                \mathbf{w} \right)\right)}\\ =
\avrg{
 -f \left(\pd{\epsilon}{x_i}+
     (\mathbf{p}-m\mathbf{w})\pd{\mathbf{w}}{x_i}\right)
 -f p_i \left(\frac{\partial^2 \epsilon}{\partial \mathbf{r} \partial \mathbf{p}}+
        \pd{\mathbf{w}}{\mathbf{r}}\right)
 -p_i \pd{f}{\mathbf{r}}
      \pd{\epsilon}{\mathbf{p}}
 -p_i \pd{f}{\mathbf{r}} \mathbf{w}}\\ =
  - \avrg{f \pd{\epsilon}{x_i}+
         f p_i \frac{\partial^2 \epsilon}{\partial \mathbf{r} \partial \mathbf{p}}+
         p_i \pd{f}{\mathbf{r}}
             \pd{\epsilon}{\mathbf{p}}}
 + \avrg{f} m \mathbf{w} \pd{\mathbf{w}}{x_i}
 - \avrg{  \mathbf{p}   \pd{\mathbf{w}}{x_i} f
        +  p_i \pd{\mathbf{w}}{\mathbf{r}}   f
        +  p_i \pd{f}{\mathbf{r}}   \mathbf{w}}
.
\end{multline*}
The first term can be transformed as
\begin{multline*}
\avrg{f \pd{\epsilon}{x_i}+
         f p_i \frac{\partial^2 \epsilon}{\partial \mathbf{r} \partial \mathbf{p}}+
         p_i \pd{f}{\mathbf{r}}
             \pd{\epsilon}{\mathbf{p}}}=
\avrg{  f p_i \frac{\partial^2 \epsilon}{\partial \mathbf{r} \partial
\mathbf{p}}+
        f     \pd{\epsilon}{x_i}+
        p_i \pd{f}{\mathbf{r}}
        \left(T \pd{ \Z }{\mathbf{p}} + \mathbf{v}\right)}\\ =
\avrg{  f p_i \frac{\partial^2 \epsilon}{\partial \mathbf{r} \partial
\mathbf{p}}+
        f     \pd{\epsilon}{x_i} +
        p_i \pd{f}{\mathbf{r}} \mathbf{v} +
        p_i \pd{f}{ \Z } \pd{ \Z }{\mathbf{p}}
            \left(\pd{\epsilon}{\mathbf{r}} -
                 p_k \pd{v_k}{\mathbf{r}} -
                 m\pd{\phi}{\mathbf{r}} -
                  \Z \pd{T}{\mathbf{r}} \right)
         }\\ =
\avrg{  p_i \pd{f}{\mathbf{r}} \mathbf{v} +
        m f \pd{\phi}{x_i} +
        f \mathbf{p}\pd{\mathbf{v}}{x_i} +
        f  \Z \pd{T}{x_i} +
        p_i f\pd{\mathbf{v}}{\mathbf{r}}+
        p_i f\pd{\Z}{\mathbf{p}}\pd{T}{\mathbf{r}}}\\ =
 n m \pd{\phi}{x_i} +
 \mathbf{P} \pd{\mathbf{v}}{x_i} +
 \pd{}{\mathbf{r}}(P_i\mathbf{v})+
 \avrg{\delta_{ik} f  \Z  +
              p_i f\pd{ \Z }{p_k}
      }\pd{T}{x_k}\\ =
 n m \pd{\phi}{x_i} +
 \mathbf{P} \pd{\mathbf{v}}{x_i} +
 \pd{}{\mathbf{r}}(P_i\mathbf{v})+
 \avrg{\delta_{ik} f \ln\frac{1+f}{f} -
              p_i \pd{ \Z }{p_k} \pd{}{ \Z }\ln(f+1)
      }\pd{T}{x_k}\\ =
 n m \pd{\phi}{x_i} +
 \mathbf{P} \pd{\mathbf{v}}{x_i} +
 \pd{}{\mathbf{r}}(P_i\mathbf{v})+
 \avrg{f \ln\frac{1+f}{f} + \ln(f+1)}\pd{T}{x_i}\\ =
 n m \pd{\phi}{x_i} +
 \mathbf{P} \pd{\mathbf{v}}{x_i} +
 \pd{}{\mathbf{r}}(P_i\mathbf{v})+
 S\pd{T}{x_i}.
\end{multline*}
Consequently,
\begin{multline}
\label{dotP}
 \dot{P_i}=
   n m \mathbf{w} \pd{\mathbf{w}}{x_i}
 - \mathbf{P} \pd{\mathbf{w}}{x_i}
 - \pd{}{\mathbf{r}} P_i \mathbf{w}
 - n m \pd{\phi}{x_i}
 - \mathbf{P} \pd{\mathbf{v}}{x_i}
 - \pd{}{\mathbf{r}}(P_i\mathbf{v})
 - S\pd{T}{x_i}
\\ =
n m \pd{}{x_i}\left(\frac{w^2}{2} - \phi\right)
 - S\pd{T}{x_i}
 - \mathbf{P} \pd{}{x_i}( \mathbf{w} + \mathbf{v})
 - \pd{}{\mathbf{r}}(P_i (\mathbf{w} + \mathbf{v})).
\end{multline}
The desirable complete set of the two-velocity elasticity theory equations
consists of Eqs.~\eqref{dotP}, \eqref{dotE} (with $Q$ defined
by~\eqref{energy_flux}), and~\eqref{dotJ} (with $\Pi_{ij}$ defined
by~\eqref{momentum_flux}).

\section{Boundary Conditions}
\label{section_bound_cond}
 We now turn to boundary conditions. They
immediately follow from the conservation relations to be satisfied at the
interface. It is much easier to perform all transformations in the frame of
reference of the interface itself. All the velocities are therefore taken
relative to the boundary. Moreover, we simplify the problem by restricting it
to the one-dimensional case: all fluxes are supposed to be perpendicular to
the flat surface; we let the $z$ axis run along this direction. Since no
curvature is ascribed to the surface, we ignore capillary effects. All
calculations done here are valid within the linear approximation. Naturally,
boundary conditions should depend on the type of the boundary and on the type
of the media on the other side of the interface. We begin with the situation
extensively discussed in literature, the solid--liquid
interface~\cite{nozieres}. Because the possibility of the mass flux through
the lattice is taken into account, the results are different, however.

\subsection{Solid--Liquid Interface}
\begin{figure}[h]
\begin{center}
\includegraphics[scale=1]{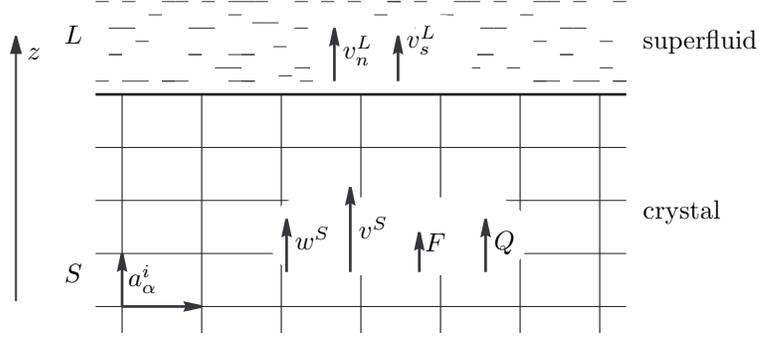}
\end{center}
\caption{Solid--liquid boundary: fluxes in one dimension}\label{sl}
\end{figure}
The liquid on the other side of the interface (being superfluid) is
characterized by the chemical potential $\mu^L$, normal and superfluid
densities $\rho^L_n$ and $\rho^L_s$, normal and superfluid velocities $v^L_n$
and $v^L_s$, temperature $T^L$, pressure $p^L$, and the entropy density $S^L$
(see Fig.~\ref{sl}).
\begin{equation}
\label{sl_set}
 \left\{
 \begin{aligned}
 R + S^S (v^S + w^S)
 &= S^L v^L_n,\\
 w^S (E^S-E^S_l+\Lambda^S_{zz}) + v^S(T^S S^S + m^S n^S
    \phi^S)
 &= \mu^L (\rho^L_s v^L_s + \rho^L_n v^L_n) + S^L T^L v^L_n,\\
 v^S m^S n^S + \rho^S w^S
 &= \rho^L_s v^L_s + \rho^L_n v^L_n,\\
 \Lambda^S_{zz} - E^S_l &= p^L.
 \end{aligned}
\right.
\end{equation}
The superscript $S$ indicates that the appropriate quantities refer to the
solid. The first equation is the entropy growth condition, where $R$ is the
surface dissipative function. The last three equations in~\eqref{sl_set} are
simply the requirements for the energy, mass, and momentum conservation for
the surface, respectively. The surface dissipative function must be a
positive square form. Using~\eqref{sl_set} and~\eqref{lambda_ij}, it can be
expressed as
\begin{multline*}
R T^L =
  v^S \left( S^S ( T^S-T^L) + m^S n^S (\phi^S - \mu^L) \right)
  +
  w^S \left( E^S-E^S_l +\Lambda^S_{zz} -\rho^S\mu^L - T^L S^S \right)\\
=
  v^S \left( m^S n^S (\phi^S - \mu^L) + S^S ( T^S-T^L) \right)
  +
  w^S \left( \lambda^S_{zz} + \rho^S(\phi^S-\mu^L) + S^S (T^S - T^L)\right)
 .
\end{multline*}

We now recall that the solid--liquid boundary can be either atomically-rough
or atomically-smooth, depending on the temperature. The nature of the surface
may (or may not) impose certain restrictions on the dynamics. For both types
of the surface, the equation
\begin{equation}
\label{p_is_p}
 \Lambda^S_{zz} - E^S_l = p^L
\end{equation}
is satisfied.

\subsubsection{Rough Surface}
\label{section_sl_rough} Employing the Onsager principle, we obtain
\begin{equation}
\label{sl_rough}
 \begin{aligned}
 v^S &= \alpha \left( m^S n^S (\phi^S - \mu^L) + S^S ( T^S-T^L)\right)
     + \eta   \left( \lambda^S_{zz} + \rho^S(\phi^S-\mu^L) + S^S (T^S - T^L) \right)\\
 w^S &= \eta \left( m^S n^S (\phi^S - \mu^L) + S^S ( T^S-T^L)\right)
     + \nu   \left( \lambda^S_{zz} + \rho^S(\phi^S-\mu^L) + S^S (T^S - T^L) \right)
.
 \end{aligned}
\end{equation}
The kinetic matrix
 $\begin{pmatrix}
 \alpha &\eta \\
 \eta   &\nu
 \end{pmatrix}$
is positively definite.

\subsubsection{Smooth Surface}
\label{section_sl_smooth} A smooth surface implies immobility of the
interface relative to the lattice. That is,
\[
w^S = 0.
\]
For the quasiparticle gas velocity, we then obtain a restricted version
of~\eqref{sl_rough},
\begin{equation}
\label{sl_v_smooth}
 v^S = \alpha \left( m^S n^S (\phi^S - \mu^L) + S^S ( T^S-T^L)
\right),
\end{equation}
with the kinetic coefficient $\alpha>0$.

\subsection{Solid--Wall Boundary}
\begin{figure}[h]
\begin{center}
\includegraphics[scale=1]{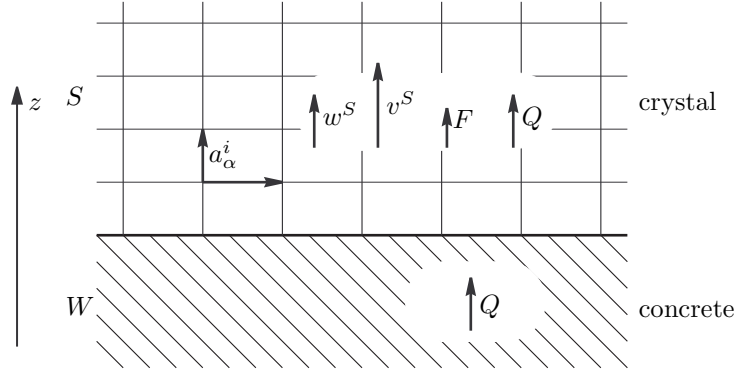}
\end{center}
\caption{Solid--wall boundary: fluxes in one dimension}\label{sw}
\end{figure}
By a wall, we imply a macroscopically flat structureless medium, in short,
``concrete''. The ``Solid--wall'' boundary occurs between solid helium and
some normal rigid solid (silica in experiment~\cite{exp}). A ``concrete''
wall is characterized by no mass flux in it (\ie through the interface). The
wall can supply an arbitrary energy flux; we let $Q$ denote the flux and
$T^W$ the wall temperature (see Fig.~\ref{sw}). Concrete is characterized by
fewer variables than liquid, and the appropriate equations are therefore
somewhat simpler.

Just like for the solid--liquid interface, the actual boundary conditions
must depend on the microscopic pattern of the surface. One can easily imagine
a smooth basal plane of the crystal adjacent to an atomically flat concrete
wall. This plane must stay at rest with respect to the wall since its motion
presupposes the creation of new atomic layer nuclei. The plane is similar to
the smooth solid--liquid interface, and we can naturally say that such an
interface is smooth. The boundary condition is then given by $w^S=v^S=0$.

Another, much more interesting scenario is realized if the interface is
slightly tilted with respect to the basal plane. Such planes may move by
growing additional nodes at the edge. This means that no restrictions are
imposed on the lattice velocity near the interface. In other words, vacancies
are allowed to freely appear and vanish on the surface (in this sense, the
surface is similar to a grid of dislocations arranged at the boundary of the
crystal that serve as sources or sinks for vacancies; similar speculations
may be found in~\cite{iml} in explaining polycrystal plasticity). We call
this type of the interface ``rough''. In this sense, the solid--wall boundary
can be either smooth or rough. The suggested growth mechanism can be applied
only to the rough boundary.

\subsubsection{Rough Boundary}
 \label{section_sw_rough}
Assuming the boundary to be rough and using the same approach as for the
liquid, we write the conservation laws
\[
\left\{
 \begin{aligned}
 S^S (v^S + w^S)
 &= R + Q/T^W,\\
 w^S (E^S-E^S_l+\Lambda^S_{zz}) + v^S(T^S S^S + m^S n^S
    \phi^S)
 &= Q,\\
 v^S m^S n^S + \rho^S w^S
 &= 0.
 \end{aligned}
\right.
\]
Again, $R$ is the surface dissipative function,
\begin{multline*}
R T^W =
 v^S\left(
            \frac{m^S n^S}{\rho^S} (E^S-E^S_l+\Lambda^S_{zz}
            - \rho^S \phi^S - T^W S^S)
            + S^S (T^W - T^S )
       \right)\\
=
 v^S\left(
            \frac{m^S n^S}{\rho^S} \lambda^S_{zz}
            + \left(1-\frac{m^S n^S}{\rho^S}\right) S^S (T^W - T^S )
       \right)
  .
\end{multline*}
It must be positive, and for the quasiparticle velocity on the surface we
therefore have
\begin{equation}
\label{sw_v_rough} v^S =
 \beta \left(
            \frac{m^S n^S}{\rho^S} \lambda^S_{zz}
            + \left(1-\frac{m^S n^S}{\rho^S}\right) S^S (T^W - T^S )
       \right),
\end{equation}
where $\beta > 0$ is the surface kinetic coefficient.

\section{The Growth Rate}
\label{section_growth}
We now use the equations and boundary conditions obtained above. The physical
system discussed in what follows is solid helium with elementary excitations
represented by phonons and vacancies. We first introduce a certain amount of
friction between the quasiparticle gas and the lattice. To obtain a
physically sound result, we again restrict our analysis to one dimension.
Furthermore, for simplicity, all our calculations are performed within the
linear approximation. We can write the quasimomentum density (with the
superscript $S$ omitted) as $K=\rho_K v$, where
\begin{equation}
\label{K_approx} \rho_K \sim \frac{\hbar}{a^4}\frac{T^4}{\Theta_D^4 c} .
\end{equation}
Here, $\Theta_D$ is the Debye temperature, $c$ is the velocity of sound in
the crystal, and $a$ is the lattice period. The last equation is quite
obvious. It follows from the fact that in the low-temperature region, the
quasimomentum is mainly associated with phonons (the number of vacancies is
exponentially small). The result therefore coincides with the one for the
mass density (and the momentum density) of the normal component of the
superfluid, $\rho_K \sim \rho_n \sim T^4/\hbar^3 c^5$ (see \cite{LL9}).

\begin{figure}[h]
\begin{center}
\includegraphics[scale=1]{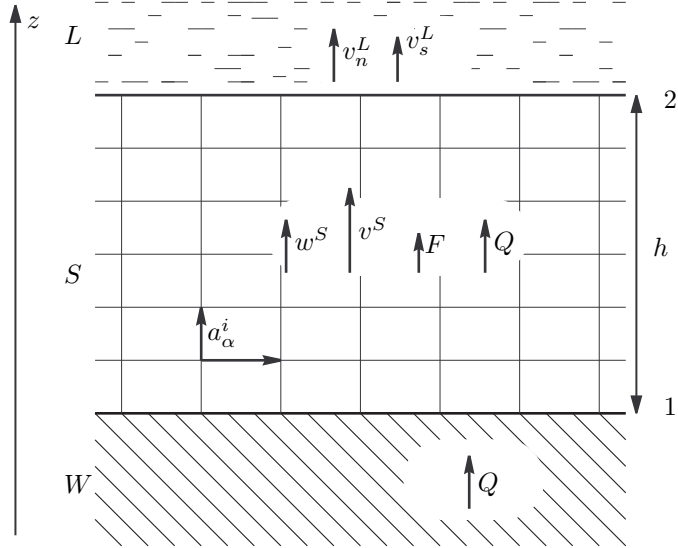}
\end{center}
\caption{Crystal growth in one dimension} \label{wsl}
\end{figure}

To describe (rare) Umklapp events, we introduce the appropriate relaxation
time parameter $\tau_U$. It is a ``between-Umklapp-collision time''.
From~\eqref{dotP}, we then have
 \[
 0=\dot{K}= -m n \nabla \phi - S \nabla T -
\frac{K}{\tau_U}
= -m n \nabla \phi- S \nabla T - \rho_K \frac{v}{\tau_U}.
\]
It is worth mentioning that $\tau_U$ may well depend on both phonons and
vacancies, despite the fact that the population of vacancies is far lower
than that of phonons. For instance, if the vacancy energy band is
sufficiently narrow, then the probability of Umklapp processes is
significantly higher for vacancies than for phonons. This might overcome the
low concentration of vacancies.

Interestingly enough, these formulas allow us to obtain the growth rate for a
smooth surface. The quasiparticles playing the crucial role here (that of
mass carriers) are vacancies, with their mass given by $m=-m_{\text{He}}$.

To estimate the rate, we write the temperature gradient as $\nabla
T=(T_2-T_1)/h$, where the subscripts 1 and 2 stand for the solid--wall and
the solid--liquid interfaces respectively (see Fig.~\ref{wsl}). Likewise, for
the chemical potential we write $\nabla \phi=(\phi_2-\phi_1)/h$. We now use
the boundary conditions~\eqref{sl_v_smooth} and~\eqref{sw_v_rough},
\begin{multline*}
(\alpha+\beta)v  =
     \alpha\beta
      \left( m n (\phi_2 - \mu^L)
            + S (T_2-T^L)
            + \frac{m n}{\rho} \lambda_{zz1}
            + \left(1-\frac{m n}{\rho}\right) S (T^W - T_1 )
      \right)
\\ =
     \alpha\beta
      \biggl\{ m n (\phi_2 - \phi_1 )
            + S (T_2 - T_1)
            + S (T^W - T^L )
            + m n \left( \phi_1 - \mu^L
            + \frac{1}{\rho} \left(\lambda_{zz1}
            - S (T^W - T_1 )\right)
            \right)
      \biggr\}
\\ =
     \alpha\beta
      \biggl\{ - \rho_K h \frac{v}{\tau_U}
            + S (T^W - T^L )
            + m n \left( \frac{1}{\rho} \left(\lambda_{zz1}
            - S (T^W - T_1 )\right) + \phi_1
            - \mu^L
            \right)
      \biggr\}
.
\end{multline*}
In other words,
\[
 v\left(\frac{1}{\alpha}+\frac{1}{\beta}
  +\frac{\rho_K h}{\tau_U}
   \right)=
            S (T^W - T^L )
            + m n \left( \frac{1}{\rho} \left(\lambda_{zz1}
            - S (T^W - T_1 )\right) + \phi_1
            - \mu^L
            \right),
\]
where (using Eqs.~\eqref{lambda_ij} and~\eqref{p_is_p})
\[
\lambda_{zz1}= \Lambda_{zz1} - T_1 S_1  -E_{l1} + E_1 - \phi_1 \rho_1 =
 p^L - T_1 S_1 + E_1 - \phi_1 \rho_1.
\]
In equilibrium, $\lambda_{zz1}=0$, $\phi_1 = \mu^L$, and $T^W=T^L=T_1$. If
the liquid temperature and pressure change by $\Delta T$ and $\Delta p$, we
can write an equation for the growth rate~$v_g$,
\begin{multline}
\label{growth_rate_general}
 -v_g \frac{\rho}{m n} \left(\frac{1}{\alpha}+\frac{1}{\beta}
  +\frac{\rho_K h}{\tau_U}
   \right)=
            - S \Delta T\\
            + m n \left( \frac{1}{\rho} \left(
 \Delta p - S \Delta T_1  - T \Delta S_1 +
 \Delta E_1 - \rho \Delta \phi_1  - \phi \Delta \rho_1
            + S \Delta T_1 \right) + \Delta \phi_1
            + \frac{S^L\Delta T -\Delta p }{ \rho^L}
            \right)\\
=
            - S \Delta T
            + m n \left( \frac{\Delta p}{\rho}
            + \frac{S^L\Delta T -\Delta p }{\rho^L}
            \right)
=
            - \left(S - S^L\frac{m n}{\rho^L}\right) \Delta T
            + m n  \left( \frac{1}{\rho}
            - \frac{1}{\rho^L}
            \right)\Delta p
,
\end{multline}
where we used Eq.~\eqref{E0_macroscopic}, the thermodynamic equality
 $\D \mu = (\D p -S\D T )/ \rho$
for the liquid, and the obvious relation
 $v_g=-v m n / \rho$.

We now consider this equation with the second term in the right-hand side
equal to zero. This is a usual scenario for heat conductivity measurements.
The heat flux $Q = v T S$ can then be expressed as
\[
Q= -\frac{ \Delta T}{R_{K1}+R_{K2}+h/\kappa},
\]
where $R_{K1}$ and $R_{K2}$ are the Kapitza thermal resistances on the
solid--wall and solid--liquid boundaries and $\kappa$ is the heat
conductivity of the crystal. Taking the inequalities $R_{K2}\ll R_{K1}$, $m n
\ll \rho$, and $\rho-\rho^L \ll \rho$ into account, we immediately obtain
\[
R_{K1}= \frac{1}{\beta T S^2}, \quad \kappa = \frac{\tau_U T S^2}{\rho_K}.
\]
As a result, the growth rate is given by
\[
v_g=\frac{mn}{\rho(T S^2 R_{K}+ T S^2 h/\kappa)}
            \left(
S \Delta T + m n \frac{\rho-\rho^L}{\rho^2}\Delta p
            \right)
            .
\]
Strictly speaking, the last equality implies that thermodynamic properties of
the crystal mainly depend on phonons, while the contribution of vacancies to
the effect under consideration is limited to the mass transfer. The growth
rate here depends on the overpressure as well as on the temperature
difference between the liquid and the wall.

In the real experiment~\cite{exp}, the temperature is lower than 100\,mK. In
this region, the phonon free path is much longer than the experimental cell
size and the impact of phonons on the vacancy behavior is proportional to a
high power of the small ratio $T/\Theta_D$. Consequently, as $\Theta_D
\rightarrow \infty$, we can consider the vacancy gas as an independent
component and neglect the influence of phonons on it. The crystal growth is
accounted for by the presence of vacancies; hence, to estimate the growth
rate in the experimental situation, we can simply substitute the vacancy-only
quantities for all variables in Eq.~\eqref{growth_rate_general}. Since it
seems that there were essentially no temperature gradients in the experiment,
we consider the isothermal case $\Delta T=0$.

The kinetic coefficients $\alpha$ and $\beta$ are determined by the vacancy
annihilation probabilities in vacancy--surface collisions. The vacancy gas
velocity $v$ at the interface should be expressed in terms of the
accommodation coefficient $W$ (which is the ratio of the number of
annihilated vacancies to the total number of incident vacancies) as
\begin{equation}
\label{sw_v_microscopic}
 v \sim \frac{\Delta f}{f} V_T W,
\end{equation}
where $\Delta f \sim f m \Delta \mu /T $ is the difference between the
incident and reflected distribution functions and $V_T\sim \sqrt{T/m^*}$ is
the thermal velocity. Here, $m^*$ is the effective mass near the bottom of
the vacancy energy band. This mass can be estimated from the energy band
width $\Delta$ as $m^*\sim \hbar^2 /(a^2\Delta)$.

The accommodation coefficient $W$, like any other inelastic process
probability in quasiparticle--surface interactions~\cite{accomodation}, is
approximately the squared ratio of the lattice period to de Broglie
wavelength,
\[
W\sim \left(\frac{a}{\lambda}\right)^2 \sim \frac{T}{\Delta}.
\]
We can now compare Eqs.~\eqref{sw_v_microscopic}, \eqref{sl_v_smooth},
and~\eqref{sw_v_rough}. For the coefficients, this yields,
\[
\alpha \sim \beta \sim \frac{a}{n\hbar}\sqrt{\frac{T}{\Delta}}.
\]

An estimate of the relaxation time $\tau_N$ characterizing the normal
(non-Umklapp) vacancy collisions can be obtained from
\[
\tau_N \sim \frac{1}{n \sigma V_T},
\]
where $\sigma\sim a^2$ is the vacancy--vacancy scattering cross-section. The
Umklapp relaxation time is exponentially longer $\tau_U \sim \tau_N \exp
(\Delta_U / T)$, where $\Delta_U < \Delta$ is a certain energy specific to
the vacancy Umklapp processes.

Using the obvious relation $\rho_K = m^* n$, we proceed  to the growth rate.
From Eq.~\eqref{growth_rate_general}, it follows that
\begin{multline*}
 v_g \sim
  \frac{m^2 n^2(\rho-\rho^L)}{(1/\alpha+1/\beta+m^* n/\tau_U)\rho^3}\Delta p
     \sim
  \frac{m^2 n^2(\rho-\rho^L)}{\left(
  \sqrt{\Delta/T} n\hbar/a
   + h m^* n^2 a^2 V_T\exp
(-\Delta_U / T) \right)\rho^3}\Delta p \\
\sim \frac{a^4 (\rho-\rho^L) }{\hbar \rho}
  \left(
  \sqrt{\frac{\Delta}{T}}e^{\epsilon_0 / T}
   + \frac{h}{a}\sqrt{\frac{T}{\Delta}}e^{-\Delta_U / T}
   \right)^{-1}\Delta p
.
\end{multline*}
Here, as an estimate, we set $\rho a^3 \sim m$ and $na^3\sim \exp(-\epsilon_0
/ T)$, where $\epsilon_0$ is the bottom of the vacancy energy band. For the
facet mobility $\mu^*_f=v_g/\Delta p$ introduced in~\cite{exp}, we have
\begin{equation}
\label{mobility} \mu^*_f\sim \frac{a^4}{\hbar }\frac{\rho-\rho^L}{\rho}
  \left(
  \sqrt{\frac{\Delta}{T}} e^{\epsilon_0 / T}
   + \frac{h}{a}\sqrt{\frac{T}{\Delta}} e^{-\Delta_U / T}
   \right)^{-1}.
\end{equation}

\section{Conclusion}
Formula~\eqref{mobility} provides a reasonable correspondence between the
theory proposed here and the experiment~\cite{exp}. It suggests three main
predictions to be verified in further experiments.
\begin{itemize}
\item
The facet mobility has a maximum at some finite temperature. Should the
temperature decrease below the point of the maximum, the growth rate will
also decrease. Otherwise, if the mobility does not tend to zero as the
temperature tends to zero, this should be considered as an indication of the
presence of zero-point vacancies (see~\cite{zv}).
\item
The observed growth rate depends on the height of the sample.
\item
The crystal grows at the boundary between the solid and the wall. This fact
can potentially be observed experimentally using some small foreign object
frozen into the crystal in its upper part.
\end{itemize}

\section{Acknowledgments}
This work is done under partial support from RFBR (grant 00-02-175470), INTAS
(grant 97-731), and NWO (grant 047.008.015).  We also thank K.O.Keshishev and
A.Ya.Parshin for fruitful discussions.

\end {document}